\documentclass[a4paper,twocolumn]{esapub2005} % European paper

\usepackage{graphicx}

    \def\IJMP{{\it Int. J. Mod. Phys.} }

   \def\PRL{{\it Phys. Rev. Lett.} }

 \def\frac#1#2{{\textstyle{{#1}\over
{#2}}}} 
\def\lsim{\mathrel{\rlap{\lower4pt\hbox{\hskip1pt$\sim$}}
\raise1pt\hbox{$<$}}}
\def\gsim{\mathrel{\rlap{\lower4pt\hbox{\hskip1pt$\sim$}}
\raise1pt\hbox{$>$}}} \def\sqr#1#2{{\vcenter{\vbox{\hrule
height.#2pt \hbox{\vrule width.#2pt height#1pt \kern#1pt \vrule
width.#2pt} \hrule height.#2pt}}}}

\def\beq{\begin{equation}} \def\eeq{\end{equation}}
\def\beqa{\begin{eqnarray}} \def\eeqa{\end{eqnarray}}

\long\def\symbolfootnote[#1]#2{\begingroup
\def\thefootnote{\fnsymbol{footnote}}\footnote[#1]{#2}\endgroup}

\usepackage{times}

\title{Probing the flyby anomaly with the Galileo constellation}

\author{O. Bertolami$^{1,2}$}\author{F. Francisco$^2$}\author{P.J.S. Gil$^3$}\author{J. P\'aramos$^{2,*}$}

\affil{
	$1$ - Departamento de F\'{\i}sica e Astronomia, Faculdade de Ci\^encias, Universidade do Porto, \\
	Rua do Campo Alegre 687, 4169-007 , Porto, Portugal
}

\affil{
	$2$ - Instituto de Plasmas e Fus\~ao Nuclear, Instituto Superior T\'ecnico, \\
	Av.\ Rovisco Pais 1, 1049-001 Lisboa, Portugal
}

\affil{
	$3$ - Departamento de Engenharia Mec\^anica and IDMEC - Instituto de Engenharia Mec\^anica, Instituto Superior T\'ecnico, \\
	Av.\ Rovisco Pais 1, 1049-001 Lisboa, Portugal
	}

\affil{E-mail addresses: orfeu.bertolami@fc.up.pt, frederico.francisco@ist.utl.pt, p.gil@dem.ist.utl.pt, paramos@ist.edu
}

\affil{
	$*$ Speaker
}

\begin{document}


\maketitle

\begin{abstract}
In the last few years, the so-called flyby anomaly has been widely discussed, but remains still an illusive topic. This is due to the harsh conditions experienced during an Earth flyby as well as due to the limited data available. In this work, we assess the possibility of confirming and characterizing this anomaly by resorting to the scientific capabilities of the future Galileo constellation.

\end{abstract}


%----------------------------------------------------------%
%- Body ---------------------------------------------------%
%----------------------------------------------------------%

\section{Introduction -- The flyby anomaly}

%----------------------------------------------------------%

During the past couple of decades, a few deep-space probes that used an Earth flyby have apparently displayed an unexpected velocity change after their gravitational assist. This has become known as the \emph{flyby anomaly}.

This variation in the velocity was detected in the residuals of the analysis performed to the Doppler and ranging data --- which showed the impossibility of fitting the trajectory with a single hyperbolic arc, but allowed for a separate fit of the inward and outward paths, if an unexpected velocity shift is introduced. It is highly localized at the perigee, where tracking through the Deep Space Network (DSN) is not available (with an approximate $4~{\rm h}$ gap). The spatial resolution of the available reconstructions, resulting form the $10~{\rm s}$ interval tracking, does not allow for an accurate characterization of the effect, so that no corresponding acceleration profile exists; thus, only the variation of the probes' velocity ({\it vis-\`a-vis} kinetic energy) is known.

This flyby anomaly has so far been observed in Galileo, NEAR, Rosetta, and Cassini missions \cite{Anderson2008}. A summary of the Earth flybys observed since the 1990's is shown in Table~\ref{flyby_table}. A detailed discussion of the two Galileo (1990 and 1992) and the NEAR (1998) gravity assists is presented in Ref.~\cite{Antreasian1998}, including an analysis of the three earliest flybys where the anomaly was observed; it included an estimate of the accelerations generated by different known effects, in an attempt to single out possible error sources.

It is estimated that the average acceleration associated to the flyby anomaly is of the order of $10^{-4}~{\rm m/s^2}$ \cite{Antreasian1998}. This is measured against the Earth oblateness, other solar system bodies, relativistic corrections, atmospheric drag, Earth albedo and infrared emissions, ocean tides, solar pressure, {\it etc.}.

%%%%%%%%%%%%%%%%%%%%%%%%%%%%%%%%%%%%%%%%%%%%%%%%%%%%%%%%%%%
\begin{table*}
	\begin{center}
	\caption{Summary of orbital parameters from Earth flybys during the last couple of decades, according to Ref.~\cite{Anderson2008}.}
	\begin{tabular}{c c c c c c c}
		\hline
			Mission	& Flyby	& $e$	& Perigee		& $v_\infty$		& $\Delta v_\infty$	& $\Delta v_\infty / v_\infty$ \\
					&	Date	&		& $({\rm km})$	& $({\rm km/s})$ 	& $({\rm mm/s})$ 	& $(10^{-6})$ \\
		\hline
			Galileo	& 1990	& $2.47$& $959.9$		& $8.949$			& $3.92 \pm 0.08$	& $0.438$ \\
			Galileo	& 1992	& $3.32$& $303.1$		& $8.877$			& $-4.6 \pm 1$		& $-0.518$ \\
			NEAR	& 1998	& $1.81$& $538.8$		& $6.851$			& $13.46 \pm 0.13$	& $1.96$ \\
			Cassini	& 1999	& $5.8$	& $1173$		& $16.01$			& $-2 \pm 1$		& $-0,125$ \\
			Rosetta	& 2005	& $1.327$& $1954$		& $3.863$			& $1.80 \pm 0.05$	& $0.466$ \\
			MESSENGER& 2005	& -		& $2347$		& $4.056$			& $0.02 \pm 0.01$	& $0.0049$ \\
			Rosetta	& 2007	& -		& $\sim 2400$	& -					& $\sim 0$			& - \\
			Rosetta	& 2009	& -		& $2481$		& -					& $\sim 0$			& - \\
		\label{flyby_table}
	\end{tabular}
	\end{center}
\end{table*}
%%%%%%%%%%%%%%%%%%%%%%%%%%%%%%%%%%%%%%%%%%%%%%%%%%%%%%%%%%%

%----------------------------------------------------------%

Subsequently, Ref.~\cite{Lammerzahl2006} extended this discussion to other possible error sources, comparing this $10^{-4}~{\rm m/s^2}$ figure with several possible origins for additional unaccounted for accelerations: these include the atmosphere, ocean tides, solid tides, spacecraft charging, magnetic moments, Earth albedo, solar wind and spin-rotation coupling. The authors conclude that all of the considered effects are several orders of magnitude below the flyby anomaly.

From these references, one can compile Table \ref{error_sources_table} for a quick overview of all effects. Taking into account a value of $10^{-4}~{\rm m/s^2}$ for the hypothetical flyby anomaly, all listed effects, except the Earth oblateness have lower orders of magnitude. This raises the issue of possible errors in the gravitational model of Earth: however, attempts to solve the flyby problem by changing the related second dynamic form factor $J_2$ have yielded unreasonable solutions, and are unable to account for all flybys \cite{Antreasian1998}.

%%%%%%%%%%%%%%%%%%%%%%%%%%%%%%%%%%%%%%%%%%%%%%%%%%%%%%%%%%%
\begin{table}
	\begin{center}
	\caption{List of orders of magnitude of possible error sources during Earth flybys, as discussed in Refs. \cite{Antreasian1998, Lammerzahl2006}.}
	\begin{tabular}{c c}
		\hline
			Effect				& Magnitude \\
								& $({\rm m/s^2})$ \\
		\hline
			Earth Oblateness		& $10^{-2}$	\\
			Other solar system bodies & $10^{-5}$	\\
			Relativity			& $10^{-7}$	\\
			Atmospheric drag	& $10^{-7}$	\\
			Ocean and Earth tides	& $10^{-7}$	\\
			Solar pressure		& $10^{-7}$	\\
			Earth infrared		& $10^{-7}$	\\
			Spacecraft charge	& $10^{-8}$	\\
			Earth albedo		& $10^{-9}$	\\
			Solar wind			& $10^{-9}$	\\
			Magnetic moment		& $10^{-15}$	\\
		\label{error_sources_table}
	\end{tabular}
	\end{center}
\end{table}
%%%%%%%%%%%%%%%%%%%%%%%%%%%%%%%%%%%%%%%%%%%%%%%%%%%%%%%%%%%

An empirical formula to fit the flyby parameters has been proposed as a function of the declinations of the incoming and outgoing asymptotic velocity vectors, $\delta_i$ and $\delta_o $, respectively \cite{Anderson2008}:

\begin{equation}
	{\Delta V_\infty \over V_\infty} = K (\cos \delta_i - \delta_o). \label{modelPRL}
\end{equation}
The constant $K$ is expressed in terms of the Earth's rotation velocity $\omega_E$, its radius $R_E$ and the speed of light $c$ as 

\beq K = {2 \omega_e R_e \over c}.\eeq

\noindent This identification is suggestive, as it evokes the general form of the outer metric due to a rotating body \cite{Ashby},

\begin{equation}
	ds^2 = \left(1 + 2{V - \Phi_0 \over c^2} \right)dt^2 - \left(1 - 2{V \over c^2} \right)(dr^2+d\Omega^2),
\end{equation}
with
\begin{equation}
	{\Phi_0 \over c^2} = {V_0 \over c^2} - {1 \over 2} \left( \omega_e R_e \over c \right)^2,
\end{equation}
where $V_0$ is the Newtonian potential at the equator. Following this tentative reasoning, and given the strong latitude dependence of Eq.~(\ref{modelPRL}), this expression appears to suggest that the Earth's rotation may be generating a much larger effect than the frame dragging predicted by General Relativity. This, however, is in contrast with the recent measurements of this effect performed by the Gravity Probe B probe \cite{GPB}, which orbits the Earth at a height of about $600~{\rm km}$, well within the onset zone of the reported flyby anomaly.

%----------------------------------------------------------%
%----------------------------------------------------------%

\section{Effect on GNSS systems}

In order to discuss the possible use of the available and future GNSS constellations to probe this flyby anomaly, one should first evaluate to what extent  it can affect their individual elements. Since the anomalous velocity change is only observed before and after flybys occurring at heights of the order of $1000~{\rm km}$, and the GNSS constellations are in approximately circular Medium Earth Orbits (MEO), at about $\sim 20000 ~{\rm km}$, one may empirically dismiss any effect.

One could sharpen the above argument, even though a full analysis is impossible due to the lack of spatial resolution and consequent inability to fully characterize the spatial dependence of the reported anomaly. A more elaborate discussion is found in Ref.~\cite{Lammerzahl2006}.

Notwithstanding, one takes as relevant figure of merit the anomalous acceleration $a\sim 10^{-4}~{\rm m/s^2}$, which may be assumed constant for the reasons above. In this case, Ref.~\cite{Bertolami} shows that no constant acceleration greater than $10^{-9}~{\rm m/s^2}$ can affect the GNSS constellation, since it would have otherwise been detected. Thus, one concludes that the flyby anomaly, if real, must be due to a strongly decaying force, which should drop by four orders of magnitude with a modest (about than a factor four) increase in distance, from $r = R_E + h \simeq 7000~{\rm km}$ to $r \simeq 27000~{\rm km}$. As a result, one may safely assume that the GNSS constellation is fundamentally unaffected by this putative anomaly, and may be thus employed to track probes performing gravity assists at the relevant region $h \sim 1000~{\rm km}$.

%----------------------------------------------------------%
%----------------------------------------------------------%

\section{GNSS spacecraft tracking}

The tracking of spacecraft through GNSS systems is already commercially available ({\it e.g.} EADS-Astrium's Mosaic \cite{mosaic}, NASA PiVoT \cite{pivot}). These systems are typically used to follow satellites in low earth orbit (LEO), at altitudes below those of the GNSS satellites ($h < h_{\rm GNSS} \sim 20000~{\rm km}$), where the GNSS signal is strongest.

Nevertheless, the Equator-S mission can receive front lobe signal from GPS satellites at an altitude of $61000~{\rm km}$ \cite{equator}. Furthermore, it is worth exploring the possibility of using the side and back lobes of the GPS signals \cite{Mthesis,BMthesis} to establish non-line of sight tracking and avoid the shading of the Earth. Clearly, the build up of more constellations and a rational use of multi-GNSS receivers, able to work simultaneously with different systems, will increase the accuracy of above-MEO satellite tracking.

The accuracy of GNSS spacecraft tracking is, understandably, better for lower orbits; however, it should be noted that during the apogee of highly elliptical orbits (HEO), the velocity is, of course, much slower than close to perigee. This allows for the construction of a good orbital solution, despite the decreased signal coverage \cite{Qiao,Moreau}. As a result of this trade off, the position and velocity accuracies for different types of orbit are somewhat similar, as depicted in Table~\ref{GNSS_accuracy_table}.

\begin{figure}[]
\centering
\includegraphics[width=\linewidth]{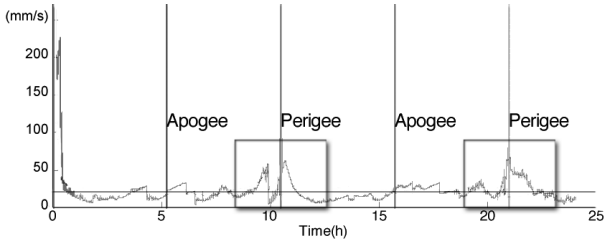}
\caption{Velocity error of multi-GNSS tracking of HEO spacecraft, adapted from Ref. \cite{Qiao}. Boxes (centered on perigee with 4~{\rm h} width) signal the gap in DNS coverage; the horizontal line corresponds to a $20~{\rm mm/s}$ accuracy.}
\label{velacc}
\end{figure}

Recall that there is no full characterization of the anomalies during the flyby, and these are detected from the mismatch between the expected and observed velocities after gravitational assist; as stated before, this is due to the inability of the DSN to track the spacecraft trajectories very close to the atmosphere, during a $\sim 4~{\rm h}$ gap. Regarding the possibility of using the GNSS in this region, Fig.~\ref{velacc} (adapted from Ref.~\cite{Qiao}) shows that, although the velocity error is maximum close to perigee, this peak is very localized: from a baseline of $\sim 20~{\rm mm/s}$ during the remaining orbit, it maxes out briefly at $\sim 100~{\rm mm/s}$ (during the first perigee approach), and converges towards $\sim 50~{\rm mm/s}$ in the subsequent perigee passings. By plotting the aforementioned gap, one sees that accuracies of $\sim 20~{\rm mm/s}$ are attainable during approximately half of this time interval.

%%%%%%%%%%%%%%%%%%%%%%%%%%%%%%%%%%%%%%%%%%%%%%%%%

%%%%%%%%%%%%%%%%%%%%%%%%%%%%%%%%%%%%%%%%%%%%%%%%%

For the study of the flyby anomaly, one would be interested in a high velocity accuracy, at least of the same order of magnitude as the observed $\Delta v_\infty \sim 1~{\rm mm/s}$. The currently available systems provide around $20~{\rm mm/s}$, which is clearly insufficient for such study. However, the presented accuracies are related to real-time orbit solutions --- which is unnecessary for the purpose of this study, and can undoubtedly be improved if offline processing is used, alongside other weak signal tracking strategies \cite{Moreau}. This, together with the increasing numbers of elements of the available (and upcoming) GNSS, lead us to conclude that it is indeed feasible to use the latter to test the flyby anomaly --- if not presently, then in the near future.

%%%%%%%%%%%%%%%%%%%%%%%%%%%%%%%%%%%%%%%%%%%%%%%%%%%%%%%%%%%
\begin{table}
	\begin{center}
	\caption{Typical accuracies expected from GNSS satellite tracking systems for LEO, MEO, Geosynchronous Earth Orbit (GEO)and HEO \cite{Qiao,Moreau,Mittnacht,Kronman}.}
	\begin{tabular}{c c c c}
		\hline
				& Apogee	& Position 	& Velocity  \\
		Orbit 	& Height  & Accuracy 		& Accuracy	\\
			&  $({\rm km})$ &  $({\rm m})$			&  $({\rm mm/s})$	\\
		\hline
			LEO		& $200$ to $2000$	& 10		& 10	\\
			MEO		& $2000$ to GEO		& 30		& 20	\\
			GEO		& $35786$			& 150		& 20	\\
			HEO		& $> 35786$			& 100		& 20	\\		
		\label{GNSS_accuracy_table}
	\end{tabular}
	\end{center}
\end{table}
%%%%%%%%%%%%%%%%%%%%%%%%%%%%%%%%%%%%%%%%%%%%%%%%%%%%%%%%%%%
 
%----------------------------------------------------------%
%----------------------------------------------------------%

\section{Probing the flyby anomaly}

We consider two options to test the flyby anomaly: an add-on to an existing mission on a Highly Elliptic Orbit (HEO), or a dedicated low-cost mission in either HEO or a hyperbolic trajectory.

In the first option, the choice would be to piggyback a GNSS receiver in an existing space mission. Since these receivers are relatively small and with reduced power consumption \cite{mosaic,pivot}, the host mission could be a small low-cost one. At apogee, a highly elliptical trajectory would present a comparable (although smaller) velocity and height as the reported anomalous gravitational assists, with the added benefit of allowing for repeated experiments.

One can regard as an example the cancelled Inner Magnetosphere Explorer (IMEX) mission of the NASA University Explorer programme, with a mass of only $ 160~{\rm kg} $ and a total budget of $ 15~M\$ $ \cite{IMEX}. The IMEX probe was to be launched as a secondary payload on a Titan IV launcher, but was cancelled due to cost overrun; it would have followed a HEO, as summarized in table \ref{tableIMEX}, which would provide a ``flyby'' velocity at perigee of $\sim 10~{\rm km/s}$, close to the reported anomalous flybys.

The more ambitious option of a dedicated mission naturally has a number of advantages over the former. One highlights the choice of orbit that can closely mimic a gravity assist --- including an hyperbolic one. However, as discussed above, a closed orbit of sufficiently high ellipticity would provide for multiple flybys, increasing the quality of the obtained data and allowing for a better characterization of the anomaly.

Furthermore, possible error sources such as aerodynamic and thermal effects close to apogee could be more closely controlled with a dedicated mission: for instance, the spacecraft could be enclosed in a spherical radio-transparent body, so to simplify modelling and reduce directional effects; if put into a spin, any accidental anisotropies would be averaged out, yielding a much cleaner testbed for the desired experiment.

This mission would require a micro-satellite with a mass under $100~{\rm kg}$ and a budget caped at less than 15 M\$, the cost of the IMEX mission. This upper bound is rather straightforward to argue by comparison: firstly, one does not anticipate any additional spending due to the simplified spherical design over the more complex IMEX probe; secondly, the scientific instrumentation found in the latter is replaced by a multi-GNSS receiver in the discussed dedicated mission, thus lowering the total cost.

As the purpose of this paper is to present the feasibility of using GNSS to probe the flyby anomaly, this estimative aims at illustrating the low cost of a dedicated mission for such purpose. Nevertheless, the actual cost could in principle be much lower than  15 M\$, not only due to the inherently simpler design and instrumentation, but also because of the ongoing trend of decreased micro-satellite costs --- reflecting advances in miniaturization, lower power consumption and improved industrial processes \cite{costs}.

%%%%%%%%%%%%%%%%%%%%%%%%%%%%%%%%%%%%%%%%%%%%%%%%%%%%%%%%%%%
\begin{table}

	\begin{center}
	\caption{Summary of orbital parameters of IMEX's HEO and a similar hyperbolic trajectory.}
	\begin{tabular}{c c c}
		\hline
									& IMEX	& Hyperbolic \\
		\hline
			Perigee					& $349~{\rm km}$ & $349~{\rm km}$	\\
			Apogee					& $335800~{\rm km}$ & --	\\
			Velocity at perigee 	& $\sim 10~{\rm km/s}$ &  $11~{\rm km/s}$	\\
			Eccentricity 			& $0.72$		& $1.04$	\\
			Orbital period 			& $10.5~{\rm h}$		&	--	\\			
		\label{tableIMEX}
	\end{tabular}
	\end{center}
\end{table}
%%%%%%%%%%%%%%%%%%%%%%%%%%%%%%%%%%%%%%%%%%%%%%%%%%%%%%%%%%%

%----------------------------------------------------------%
%----------------------------------------------------------%

\section{Conclusions}

When considering the use of Galileo (or GNSS in general) to study the flyby anomaly, one finds that most available studies deal with the tracking of spacecraft in real time, which is characterized by an insufficient velocity and position accuracy to currently detect this hypothetical phenomenon. However, since this real time accuracy is only one order of magnitude above that required (in particular, $\sim 10~{\rm mm/s}$ {\it vs.} $\sim 1~{\rm mm/s}$ in velocity), one expects that this situation could change in the short-term: a thorough exploitation of available resources could lead to a suitable tracking of spacecraft below the stated accuracy, by abandoning real time solutions and instead resorting to offline processing, use of side and back lobe tracking, amongst other weak signal tracking strategies. Crucially, the use of several GNSS at once should be paramount, due to the increased coverage gained from the different geometries.

Thus, one can safely state that there is no {\it a priori} issue in using GNSS tracking to study the reported flyby anomaly. Naturally, this availability is not sufficient, as only spacecraft equipped with a (multi-)GNSS receiver would allow for such study. In this work, we have shown that such a mission could be easily deployed, either as an add-on package to an existing hub with the required highly elliptical orbit, or via a dedicated mission.

While the first scenario would provide a cheap solution, we argue that a dedicated mission could be envisaged with a higher scientific payoff, while maintaining an overall low-cost approach.

Regardless of the actual origin of the flyby anomaly (unaccounted conventional effect, numerical procedure or, more tantalizingly, new physics), we believe it offers a low-cost opportunity for displaying the scientific possibilities opened by the GNSS era --- and Galileo in particular.

%----------------------------------------------------------%
%----------------------------------------------------------%

\section*{acknowledgments}

This work was developed in the context of the third colloquium {\it Scientific and Fundamental Aspects of the Galileo Programme}, in Copenhagen,  31$^{\rm st}$  August to $2^{\rm nd}$  September 2011. The authors thank the European Space Agency and the local organization for the hospitality displayed.

%----------------------------------------------------------%
%----------------------------------------------------------%

\end{document}